\documentclass[runningheads]{llncs}
\usepackage{subfig}
\usepackage[export]{adjustbox}
\usepackage{cite}
\usepackage{balance}

\title{PROSTATE GLAND SEGMENTATION IN HISTOLOGY IMAGES VIA RESIDUAL AND MULTI-RESOLUTION U-NET\thanks{This work was supported by the Spanish Ministry of Economy and Competitiveness through project DPI2016-77869. The Titan V used for this research was donated by the NVIDIA Corporation. Preprint accepted for publication on 21st International Conference on Intelligent Data Engineering and Automated Learning - IDEAL 2020.}}

\author{Julio Silva-Rodr\'iguez\inst{1} \and
Elena Pay\'a-Bosch\inst{2} \and
Gabriel Garc\'ia\inst{2} \and Adri\'an Colomer\inst{2} \and Valery Naranjo\inst{2}}
\authorrunning{J. Silva-Rodr\'iguez et al.}
\titlerunning{PROSTATE GLAND SEGMENTATION VIA U-NET}

\institute{
Institute of Transport and Territory, \textit{Universitat Polit\`ecnica de Val\`encia}, Spain\\
\email{jjsilva@upv.es}\\
\and
Institute of Research and Innovation in Bioengineering, \textit{Universitat Polit\`ecnica de Val\`encia}, Spain\\
}

\begin{document}

% Authors
% -------
%\name{%
%\begin{tabular}{@{}c@{}}
%Julio Silva-Rodr\'iguez$^{\star }$,  
%Elena Pay\'a-Bosch$^{\dagger}$, 
%Gabriel Garc\'ia$^{\dagger}$, 
%Adri\'an Colomer$^{\dagger}$ and
%Valery Naranjo$^{\dagger}$ 
% Acknowledgements
%\thanks{This work was supported by the Spanish Ministry of Economy and %Competitiveness through project DPI2016-77869. The Titan V used for this %research was donated by the NVIDIA Corporation.}
%\end{tabular}}

% Affiliations
% ------------
%\address{%
%\begin{tabular}{cc}
%Institute of Transport and Territory$^{\star}$ & 
%Institute of Research and Innovation in Bioengineering$^{\dagger}$  \\ 
%\textit{Universitat Polit\`ecnica de Val\`encia}, Spain & 
%\textit{Universitat Polit\`ecnica de Val\`encia}, Spain \\
%jjsilva@upv.es &
%vnaranjo@dcom.upv.es
%\end{tabular}}

% Input title
% -----------
\maketitle

%% Abstract
%-----------
\begin{abstract}

Prostate cancer is one of the most prevalent cancers worldwide. One of the key factors in reducing its mortality is based on early detection. The computer-aided diagnosis systems for this task are based on the glandular structural analysis in histology images. Hence, accurate gland detection and segmentation is crucial for a successful prediction. The methodological basis of this work is a prostate gland segmentation based on \textit{U-Net} convolutional neural network architectures modified with residual and multi-resolution blocks, trained using data augmentation techniques. The residual configuration outperforms in the test subset the previous state-of-the-art approaches in an image-level comparison, reaching an average \textit{Dice Index} of $0.77$.

\end{abstract}

%% Keywords
%-----------

%% Introduction
%--------------
\section{Introduction}
\label{sec:introduction}
% ----------------
% Motivation

Prostate cancer was the second most prevalent cancer worldwide in 2018, according to the Global Cancer Observatory \cite{who}. The final diagnosis of prostate cancer is based on the visual inspection of histological biopsies performed by expert pathologists. Morphological patterns and the distribution of glands in the tissue are analyzed and classified according to the Gleason scale \cite{gleason}. The Gleason patterns range from $3$ to $5$, inversely correlating with the degree of glandular differentiation. In recent years, the development of computer-assisted diagnostic systems has increased in order to raise the level of objectivity and support the work of pathologists.  

%The visual analysis of biopsies is a very heavy workload for physicists and is known to be subjective among experts. These limitations have generated in recent years growth in the development of computer-aided-diagnosis systems in order to increase the level of objectivity and support the task of the pathologist.

\begin{figure}[htb]
\captionsetup[subfloat]{farskip=1pt,captionskip=0.8pt}
    \centering
    
      \subfloat[\label{fig1a}]{\includegraphics[width=.235\linewidth, frame]{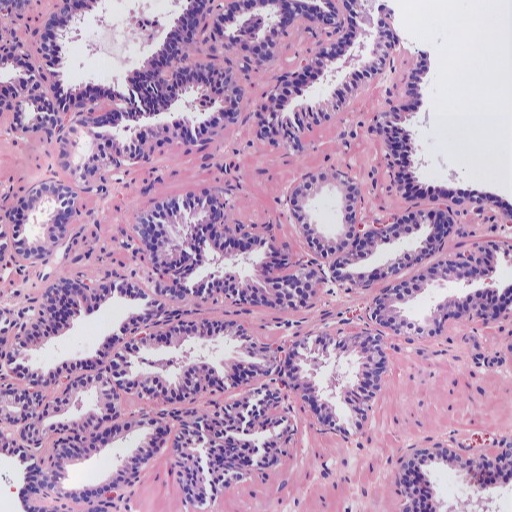}}
      \hspace*{\fill}
      \subfloat[\label{fig1b}]{\includegraphics[width=.235\linewidth, frame]{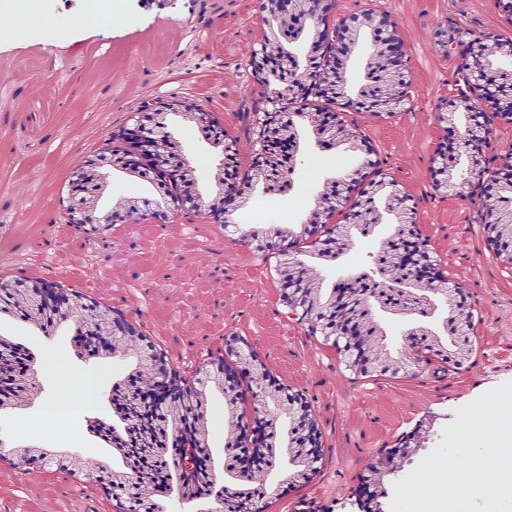}}
      \hspace*{\fill}
      \subfloat[\label{fig1c}]{\includegraphics[width=.235\linewidth, frame]{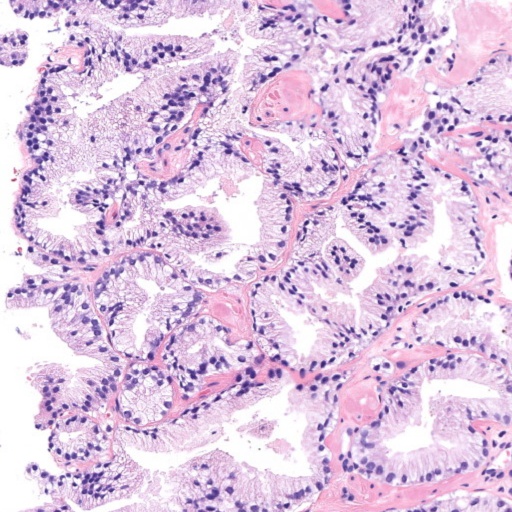}}
      \hspace*{\fill}
      \subfloat[\label{fig1d}]{\includegraphics[width=.235\linewidth, frame]{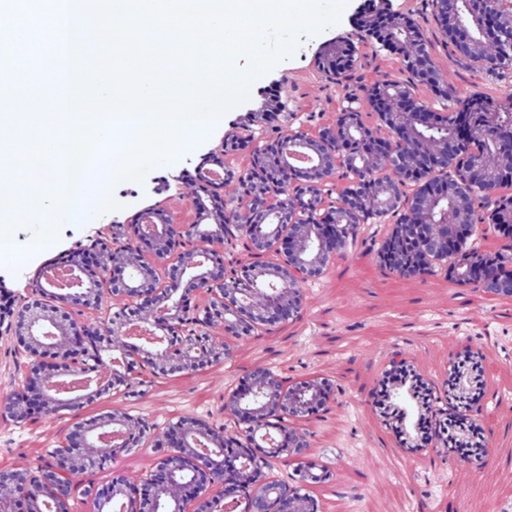}}
    \caption{Histology regions of prostate biopsies. Examples (a) and (b) present benign glands, including dilated and fusiform patterns. Images (c) and (d) contain patterns of Gleason grade $3$, with small sized and atrophic glands.}
    \label{fig1}
   
\end{figure}

One of the ways to reduce mortality in prostate cancer is through its early detection \cite{Vickers2012PredictingWhy}. For this reason, several works have focused on the first stage of prostate cancer detection by differentiating between benign and Gleason pattern $3$ glands \cite{Nguyen2012ProstateFeatures,Garcia2019First-stageLearning,Garcia2019ComputerTechniques}. The benign glands differentiate from Gleason pattern $3$ structures in the size, morphology, and density in the tissue (see Fig. \ref{fig1}). In order to automatically detect the early stage of prostate cancer, the main methodology used in the literature is based on detecting and segmenting  glands  and  then,  classifying  each individual gland. For the classification of cancerous glands, both classic approaches based on hand-driven feature extraction \cite{Nguyen2014ProstateNuclei} and modern deep-learning techniques \cite{Garcia2019First-stageLearning} have been used. Nevertheless, those results are limited by a correct detection and delimitation of the glands in the image. This encourages the development of accurate systems able to detect and segment the glandular regions.

\vspace*{-\baselineskip}

\vspace*{-\baselineskip}

% ----------------
% State of the art

For the prostate gland segmentation, different approaches have been carried out. In the work of Nguyen et al. \cite{Nguyen2012ProstateFeatures,Nguyen2014ProstateNuclei,Nguyen2010AutomatedImages,Nguyen2012StructureClassification} this procedure is based on the unsupervised clustering of the elements in the tissue, i.e. lumen, nuclei, cytoplasm, and stroma. Then, for each detected lumen, a gland is associated if enough nuclei are found in a region surrounding the lumen's contour. In the research carried out by García et al. \cite{Garcia2019First-stageLearning,Garcia2019ComputerTechniques,Garcia2018IdentificationTechniques} the components in the image are clustered by working in different color spaces, and then a Local Constrained Watershed Transform algorithm is fitted using the lumens and nuclei as internal and external markers respectively. As final step, the both aforementioned methodologies require a supervised model to differentiate between artifacts or glands. To the best of the authors' knowledge, the best performing state-of-the-art techniques for semantic segmentation, based on convolutional neural networks, have not been studied yet for prostate gland segmentation. In particular, one of the most used techniques in recent years for semantic segmentation is the \textit{U-Net} architecture, proposed for medical applications in \cite{Ronneberger2015U-net:Segmentation}.        

%Finally, a recent work of Avenel et. al \cite{Avenel2019GlandularPathology} have proposed the gland segmentation based on morphological operations over the stroma mask and the use of watershed techniques. 

% ------------------------
% Objectives and novelties

In this work, we present an \textit{U-Net}-based model that aims to segment the glandular structures in histology prostate images. To the extent of our knowledge, this is the first time that an automatic feature-learning method is used for this task. One of the main contributions of this work is an extensive validation about different convolutional block modifications and regularization techniques on the basic \textit{U-Net} architecture. The proposed convolutional block configurations are based on well-known CNN architectures of the literature (i.e. residual and Inception-based blocks). Furthermore, we study the impact of regularization approaches during the training stage based on data augmentation and using the gland contour as an independent class. Finally, we perform, as novelty, an image-level comparison of the most relevant methods in the literature for prostate gland segmentation under the same database. Using our proposed modified \textit{U-Net} with residual blocks, we outperform in the test subset previous approaches.

%% Methods
%---------

\section{Materials and methods}
\label{sec:methods}
%% Materials
%-----------

\subsection{Materials}
\label{ssec:materials}

The database used in this work consists of $47$ whole slide images (WSIs, histology prostate tissue slices digitised in high-resolution images) from $27$ different patients. The ground truth used in this work was prepared by means of a pixel-level annotation of the glandular structures in the tissue. In order to work with the high dimensional WSIs, they were sampled to $10\times$ resolution and divided into patches with size $1024^2$ and overlap of $50\%$ among them. For each image, a mask was extracted from the annotations containing the glandular tissue. The resulting database includes $982$ patches with its respective glandular masks.          

%% U-NET ARCHITECTURE FOR SEMANTIC SEGMENTATION
%-----------------------------------------------
\vspace*{-\baselineskip}
\subsection{\textit{U-Net} architecture}
\label{ssec:unet}

\begin{figure}[htb]
    \begin{center}
    \includegraphics[width=.70\textwidth]{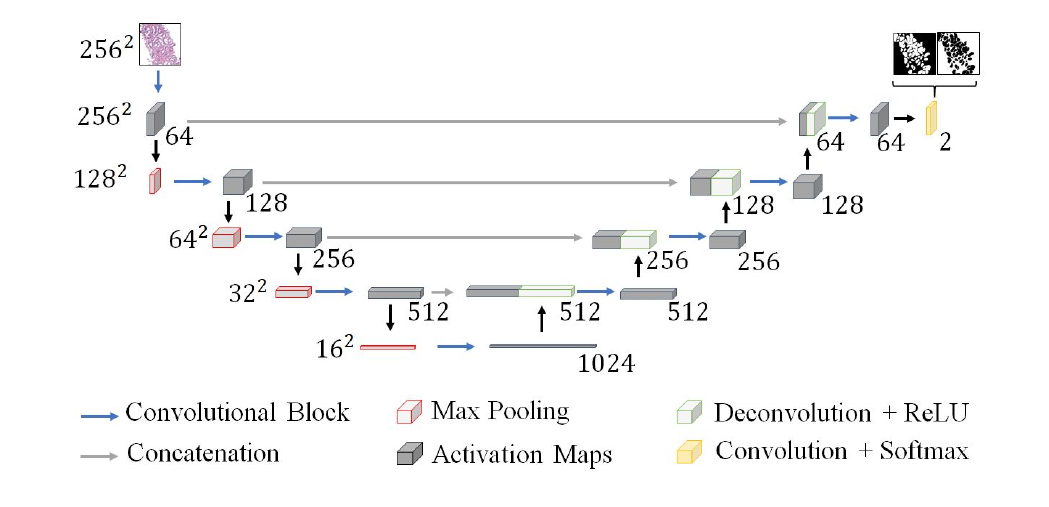}
    \caption{\textit{U-Net} architecture for prostate gland segmentation.}
    \label{fig2}
    \end{center}
\end{figure}

The gland segmentation in the prostate histology images process is carried out by means of the \textit{U-Net} convolutional neural network architecture \cite{Ronneberger2015U-net:Segmentation} (see Fig. \ref{fig2}). As input, the images of dimensions $1024^2$ are resized to $256^2$ to avoid memory problems during the training stage. The \textit{U-Net} configuration is based on a symmetric encoder-decoder path. In the encoder part, a feature extraction process is carried out based on convolutional blocks and dimensional reduction through max-pooling layers. Each block increases the number of filters in a factor of $2\times$, starting from $64$ filters up to $1024$. After each block, the max-pooling operation reduces the activation maps dimension in a factor of $2x$. The basic convolutional block (hereafter referred to as $basic$) consist of two stacked convolutional layers with filters of size $3\times3$ and ReLU activation. Then, the decoder path builds the segmentation maps, recovering the original dimensions of the image. The reconstruction process is based on deconvolutional layers with filters of size $3\times3$ and ReLU activation. These increase the spatial dimensions of the activation volume in a factor of $2\times$ while reducing the number of filters in a half. The encoder features from a specific level are joined with the resulting activation maps of the same decoder level by a concatenation operation, feeding a convolutional block that combines them. Finally, once the original image dimensions are recovered, a convolutional layer with as many filters as classes to segment and soft-max activation creates the segmentation probability maps.  

\vspace*{-\baselineskip}

%% Loss Function
%-----------------
\subsection{Loss function}
\label{ssec:dice}

The loss function defined for the training process is the categorical $Dice$. This measure takes as input the reference glands and background masks ($y$) and the predicted probability maps ($\widehat{y}$) and is defined as follows:

\begin{equation}
\label{eq:loss}
Dice(y,\widehat{y}) = \frac{1}{C}\sum_{c=1}^{C}\frac{2\sum\widehat{y}_{c}\circ y_{c}}{\sum\widehat{y}_{c}^2+y_{c}^2}
\end{equation}
%Dice(\widehat{y},y) = \frac{1}{M}\sum_{c=1}^{M}w_{c}(y_{c}log(\widehat{y}_{c})+(1-y_{c})log(1-\widehat{y_{c}}))

\noindent where $y$ is the pixel-level one hot encoding of the reference mask for each class $c$ and $\widehat{y}$ is the predicted probability map volume. %The symbol $\circ$ indicates the Hadamart product between maps.

Using a categorical average among the different classes brings robustness against class imbalance during the training process. %, avoiding underestimation of the minority class (in this case, the gland class).

%% RESIDUAL U-NET
%-----------------
\subsection{Introducing residual and multi-resolution blocks to the \textit{U-Net}}
\label{ssec:residual}

To increase the performance of the \textit{U-Net} model, different convolutional blocks are used to substitute the $basic$ configuration. In particular, residual and multi-resolution Inception-based blocks are used during the encoder and decoder stages.

The residual block \cite{He2016DeepRecognition} (from now on $RB$) is a configuration of convolutional layers that have shown good performance in deep neural networks optimisation. The residual block proposed to modify the basic U-Net consist of three convolutional layers with size $3\times3$ and ReLU activation. The first layer is in charge of normalizing the activation maps to the output' amount of filters for that block. The resultant activation maps from this layer are combined in a shortcut connection with the results of two-stacked convolutional layers via an adding operation.

%\begin{figure}[htb]
%    \begin{center}
%    \includegraphics[width=.75\textwidth]{Figures/resBlock.pdf}
%    \caption{Residual Block. $F_{in}$: number of activation maps in the input volume. $F_{out}$: %number of activation maps in the output volume. $D_{1}$, $D_{2}$: activation map dimensions.}
%    \label{fig3}
%    \end{center}
%\end{figure}

\begin{figure}[htb]
    \begin{center}
    \includegraphics[width=.60\textwidth]{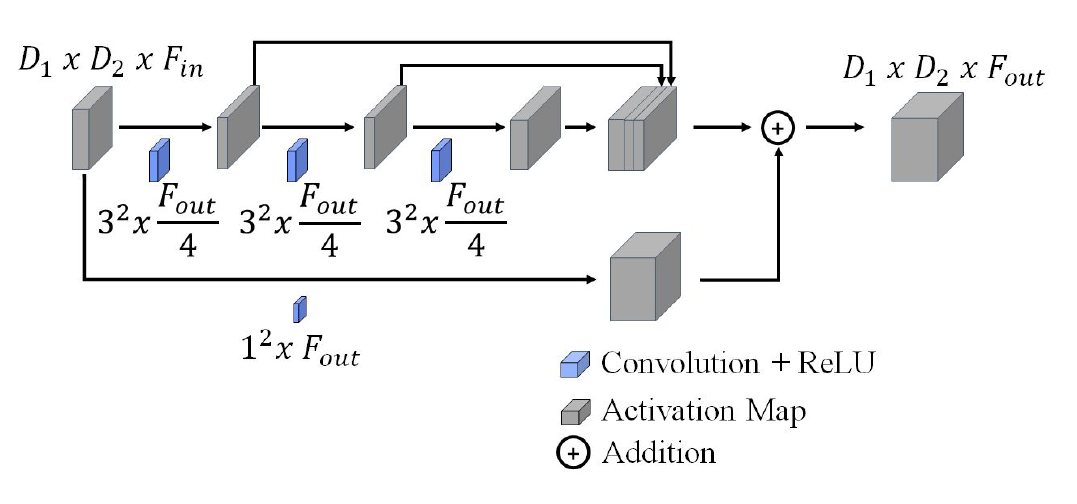}
    \caption{Multi-resolution block. $F_{in}$: activation maps in the input volume. $F_{out}$: number of activation maps in the output volume. $D$: activation map dimensions.}
    \label{fig4}
    \end{center}
\end{figure}

Regarding the multi-resolution block (referred to in this work as $MRB$), it was recently introduced in \cite{Ibtehaz2020MultiResUNet:Segmentation} as a modification of the \textit{U-Net} with gains of accuracy on different medical applications. This configuration, based on Inception blocks \cite{Szegedy2015GoingConvolutions}, combines features at different resolutions by concatenating the output of three consecutive convolutional layers of size $3\times3$ (see Fig. \ref{fig4}). The number of activation maps in the output volume ($F_{out}$) is progressively distributed in the three blocks ($\frac{F_{out}}{4}$, $\frac{F_{out}}{4}$, and $\frac{F_{out}}{2}$ respectively). Furthermore, a residual connection is established with the input activation maps, normalizing the number of maps with a convolutional layer of size $1\times1$.

\vspace*{\baselineskip}
\vspace*{-\baselineskip}

%% Regularization techniques
%---------------------------
\subsection{Regularization techniques}
\label{ssec:regularization}

To improve the training process, two regularization techniques are proposed: data augmentation and the addition of the gland border as an independent class. Data augmentation ($DA$) is applied during the training process making random translations, rotations, and mirroring are applied to the input images. Regarding the use of the gland contour as an independent class ($BC$), this strategy has shown to increase the performance in other histology image challenges such as nuclei segmentation \cite{Kumar2017APathology}. The idea is to highlight more (during the training stage) the most important region for obtaining an accurate segmentation: the boundary between the object and the background. Thus, the reference masks and the output of the \textit{U-Net} are modified with an additional class in this approach.

%% Experiments and Results
%-------------------------
\section{Experiments and Results}
\label{sec:experimental}
To validate the different \textit{U-Net} configurations and the regularization techniques, the database was divided in a patient-based $4$ groups cross-validation strategy. As figure of merit, the image-level average \textit{Dice Index} ($DI$) for both gland and background was computed. The \textit{Dice Index} for certain class $c$ is obtained from the $Dice$ function (see Equation \ref{eq:loss}) such that: $DI = 1-Dice$. The metric ranges $0$ to $1$, from null to perfect agreement.    

The different \textit{U-Net} architectures, composed of basic ($basic$), residual ($RB$) and multi-resolution ($MRB$) blocks were trained with the proposed regularisation techniques, data augmentation ($DA$) and the inclusion of the border class ($BC$), in the cross-validation groups. The training was performed in mini-batches of $8$ images, using NADAM optimiser. Regarding the learning rates, those were empirically optimised at values $5*10^{-4}$ for the $basic$ and $RB$ configurations and to $1*10^{-4}$ for the $MRB$ one. All models were trained during $250$ epochs. The results obtained in the cross-validation groups are presented in Table \ref{tab1}.

\begin{table}[htb]
\begin{center}
\caption{Results in the validation set for gland and background segmentation. The average \textit{Dice Index} is presented for the different configurations. \textit{DA}: data augmentation, \textit{BC}: border class, \textit{RB}: residual and \textit{MRB}: multi-resolution blocks.}
\label{tab1}
%\resizebox{\linewidth}{!}{
\begin{tabular}{|l|l|l|}
\hline
\multicolumn{1}{|c|}{$\mathbf{Method}$} & \multicolumn{1}{c|}{$\mathbf{DI_{gland}}$} & \multicolumn{1}{c|}{$\mathbf{DI_{background}}$}    \\ \hline\hline
$basic$                        & $0.5809(0.2377)$                               & $0.9766(0.0240)$                               \\ \hline
$basic + DA$                   & $0.6941(0.2515)$                               & $0.9845(0.01664)$                              \\ \hline
$basic + DA + BC$              & $0.6945(0.2615)$                               & $0.9842(0.0168)$                               \\ \hline\hline
$RB$                           & $0.5633(0.2340)$                               & $0.9759(0.0255)$                               \\ \hline
$RB + DA$                      & $\mathbf{0.7527(0.2075)}$                      & $\mathbf{0.9862(0.0148)}$                      \\ \hline
$RB + DA + BC$                & $0.7292(0.2395)$                               & $0.9854(0.0161)$                                \\ \hline\hline
$MRB$                         & $0.5710(0.2378)$                               & $0.9765(0.0253)$                                \\ \hline
$MRB + DA$                     & $0.7294(0.2173)$                               & $0.9843(0.0161)$                               \\ \hline
$MRB + DA + BC$                & $0.7305(0.2247)$                               & $0.9846(0.0163)$                               \\ \hline
\end{tabular}
%}
\end{center}
\end{table}

Analysing the use of different \textit{U-Net} configurations, the best performing method was the \textit{U-Net} modified with residual blocks ($RB + DA$), reaching an average $DI$ of $0.75$, and outperforming the basic architecture in $0.06$ points. Regarding the use of the multi-resolution blocks ($MRB$), an increase in the performance was not registered. Concerning the use of the gland profile as an independent class ($BC$), the results obtained were similar to the ones with just two classes.            

\begin{figure}[htb]
\captionsetup[subfloat]{farskip=1pt,captionskip=0.8pt}
    \centering
    
      \subfloat{\includegraphics[width=.235\linewidth, frame]{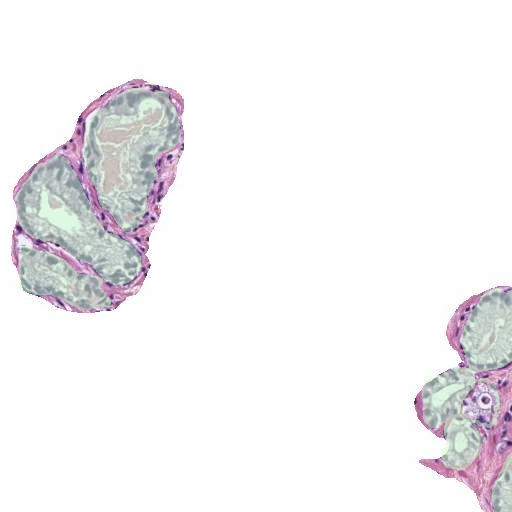}}
      \hspace*{\fill}
      \subfloat{\includegraphics[width=.235\linewidth, frame]{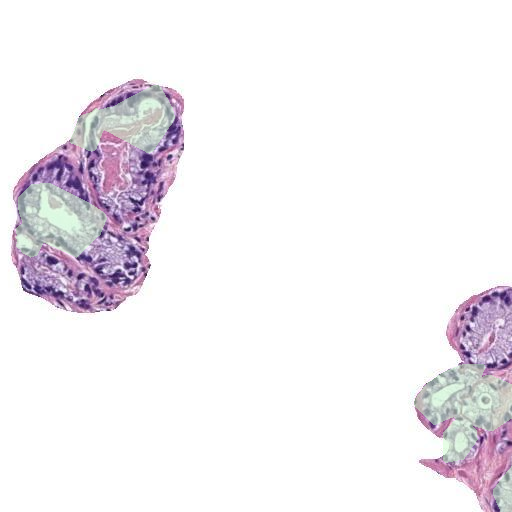}}
      \hspace*{\fill}
      \subfloat{\includegraphics[width=.235\linewidth, frame]{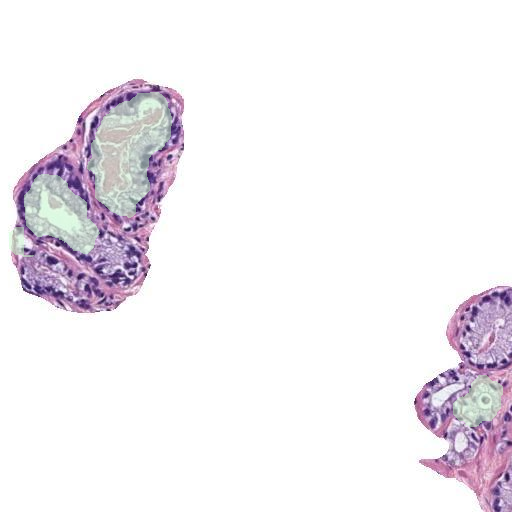}}
      \hspace*{\fill}
      \subfloat{\includegraphics[width=.235\linewidth, frame]{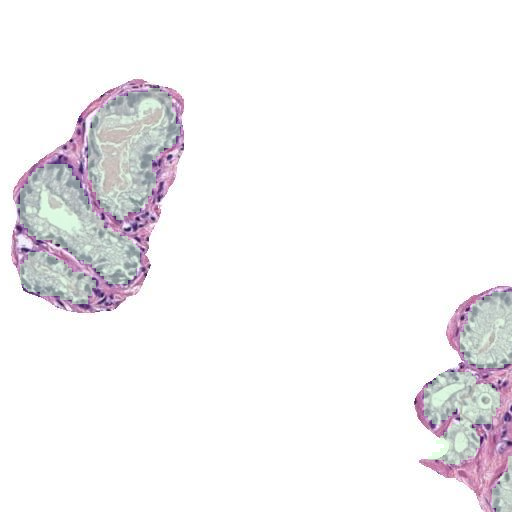}}
    
      \subfloat{\includegraphics[width=.235\linewidth, frame]{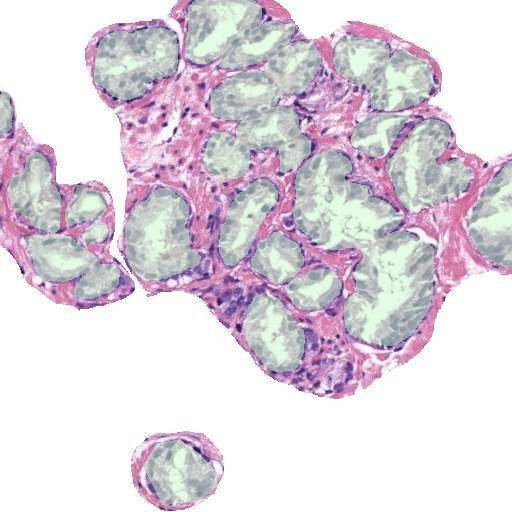}}
      \hspace*{\fill}
      \subfloat{\includegraphics[width=.235\linewidth, frame]{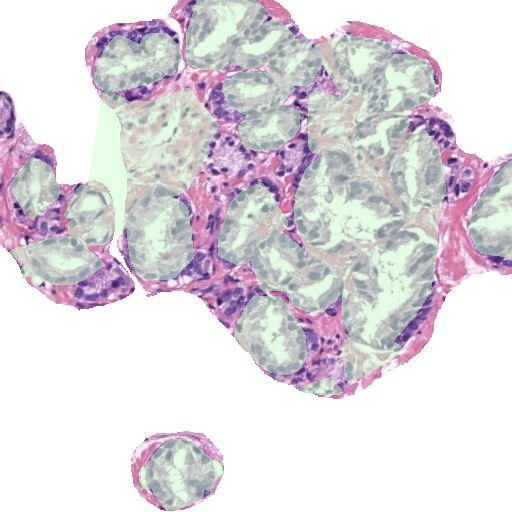}}
      \hspace*{\fill}
      \subfloat{\includegraphics[width=.235\linewidth, frame]{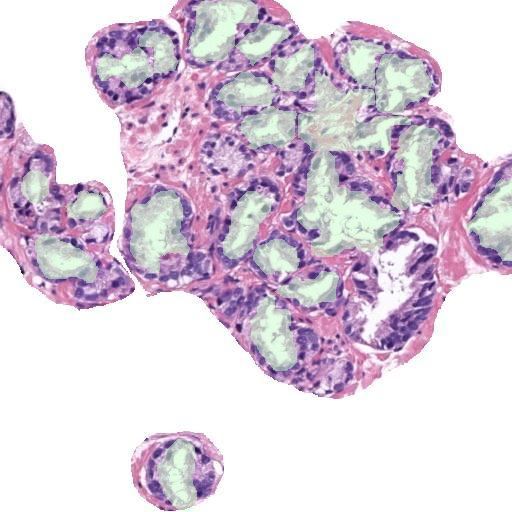}}
      \hspace*{\fill}
      \subfloat{\includegraphics[width=.235\linewidth, frame]{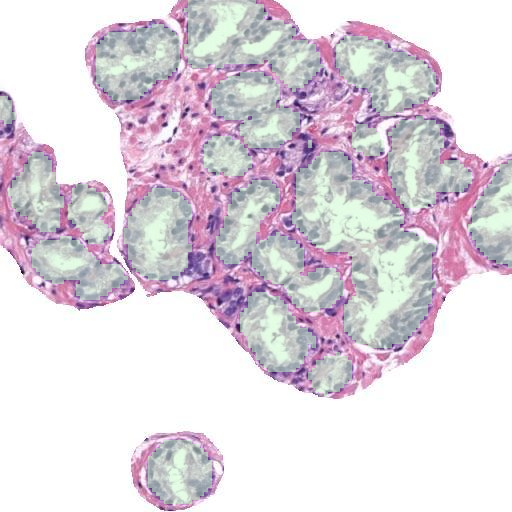}}
    
      \renewcommand{\thesubfigure}{a}
      \subfloat[\label{fig5a}]{\includegraphics[width=.235\linewidth, frame]{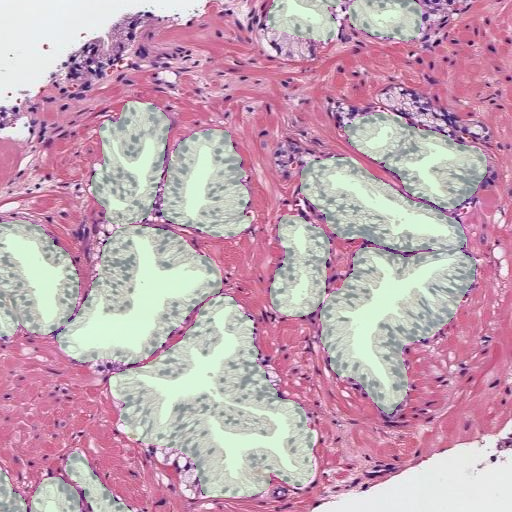}}
      \hspace*{\fill}
      \renewcommand{\thesubfigure}{b}
      \subfloat[\label{fig5b}]{\includegraphics[width=.235\linewidth, frame]{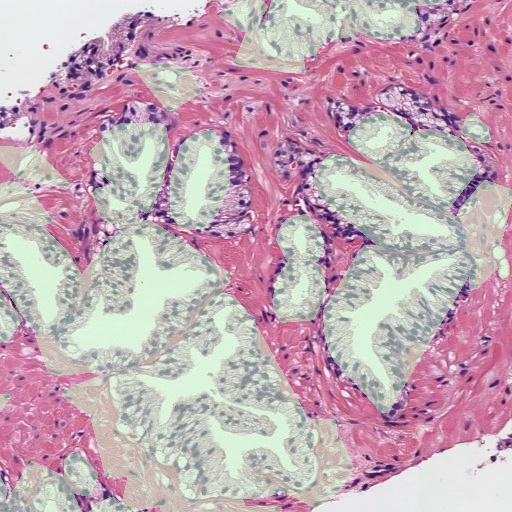}}
      \hspace*{\fill}
      \renewcommand{\thesubfigure}{c}
      \subfloat[\label{fig5c}]{\includegraphics[width=.235\linewidth, frame]{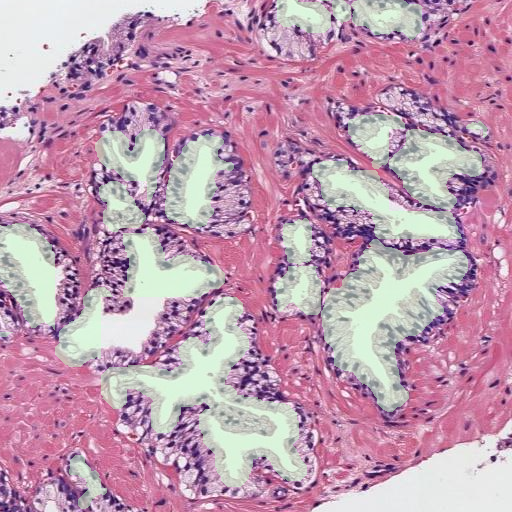}}
      \hspace*{\fill}
      \renewcommand{\thesubfigure}{d}
      \subfloat[\label{fig5d}]{\includegraphics[width=.235\linewidth, frame]{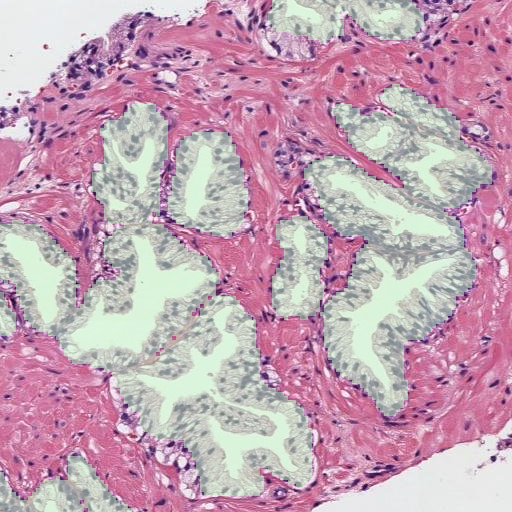}}
      
    \caption{Semantic gland segmentation in regions of images from the test set. (a): reference, (b): Nguyen et al., (c): García et al., and (d): proposed \textit{U-Net}.}
    \label{fig5}
\end{figure}

The best performing strategy, $RB + DA$, was evaluated in the test subset. A comparison of the results obtained in this cohort with the previous methods presented in the literature is challenging. While previous works report object-based metrics, we consider more important to perform an image-level comparison of the predicted segmentation maps with the ground truth, in order to take into account false negatives in object detection. For this reason, and in order to establish fair comparisons, we computed the segmentation results in our test cohort applying the main two approaches found in the literature: the work of Nguyen et al. \cite{Nguyen2012ProstateFeatures} and the research developed by García et al. \cite{Garcia2018IdentificationTechniques}. This is, to the best of the authors' knowledge, the first time in the literature that the main methods for prostate gland segmentation are compared at image level in the same database. The figures of merit obtained in the test set are presented in the Table \ref{tab2}. Representative examples of segmented images for all the approaches are shown in Fig. \ref{fig5}.
\vspace*{-\baselineskip}

%\footnote{We contacted the corresponding authors to obtain the implementation of the methods.}
%\vspace*{-\baselineskip}

\begin{table}[htb]
\begin{center}
\caption{Results in the test set for gland and background segmentation. The average \textit{Dice Index} for both classes is presented for the state-of-the-art methods and our proposed model. \textit{DA}: data augmentation and \textit{RB}: residual blocks.}
\label{tab2}
%\resizebox{\linewidth}{!}{
\begin{tabular}{|l|l|l|}
\hline
\multicolumn{1}{|c|}{$\mathbf{Method}$} & \multicolumn{1}{c|}{$\mathbf{DI_{gland}}$} & \multicolumn{1}{c|}{$\mathbf{DI_{background}}$}     \\ \hline\hline
\textit{Nguyen et al. } \cite{Nguyen2012ProstateFeatures}                & $0.5152(0.2201)$                           & $0.9661(0.0168)$                                    \\ \hline
\textit{García et al.} \cite{Garcia2018IdentificationTechniques}               & $0.5953(0.2052)$                           & $0.9845(0.01664)$                                   \\ \hline
$\textit{U-Net}+RB+DA$                     & $\mathbf{0.7708(0.2093)}$                  & $\mathbf{0.9918(0.0075)}$                           \\ \hline
\end{tabular}
%}
\end{center}
\end{table}

Our model outperformed previous methods in the test cohort, with an average $DI$ of $0.77$ for the gland class. The method proposed by Nguyen et al. and the one of García et al. obtained $0.51$ and $0.59$ respectively. The main differences were observed in glands with closed lumens (see first and second row in Fig. \ref{fig5}). The previous methods, based on lumen detection, did not segment properly those glands, while our proposed \textit{U-Net} obtains promising results. Our approach also shows a better similarity in the contour of the glands with respect the reference annotations (see third row in Fig. \ref{fig5}).

%% Conclusions
%-------------
\section{Conclusions}
\label{sec:conclusions}
In this work, we have presented modified \textit{U-Net} models with residual and multi-resolution blocks able to segment glandular structures in histology prostate images. The \textit{U-Net} with residual blocks outperforms in an image-level comparison previous approaches in the literature, reaching an average \textit{Dice Index} of $0.77$ in the test subset. Our proposed model shows better performance in both glands with closed lumens and in its shape definition. Further research will focus on studying the gains in accuracy in the first-stage cancer identification with a better gland segmentation based on our proposed \textit{U-Net}.  

%% References
%-------------
\bibliographystyle{IEEEbib}
\bibliography{refs,references}

\begin{thebibliography}{10}

\bibitem{who}
{World Health Organization},
\newblock ``Global cancer observatory,'' 2019.

\bibitem{gleason}
Donald Gleason,
\newblock ``Histologic grading of prostate cancer: A perspective, human
  pathology,'' 1992.

\bibitem{Vickers2012PredictingWhy}
Andrew~J. Vickers and Hans Lilja,
\newblock ``{Predicting prostate cancer many years before diagnosis: How and
  why?},''
\newblock {\em World Journal of Urology}, vol. 30, no. 2, pp. 131--135, 2012.

\bibitem{Nguyen2012ProstateFeatures}
Kien Nguyen, Bikash Sabata, and Anil~K. Jain,
\newblock ``{Prostate cancer grading: Gland segmentation and structural
  features},''
\newblock {\em Pattern Recognition Letters}, vol. 33, no. 7, pp. 951--961,
  2012.

\bibitem{Garcia2019First-stageLearning}
Gabriel Garc{\'{i}}a, Adrián Colomer, and Valery Naranjo,
\newblock ``{First-stage prostate cancer identification on histopathological
  images: Hand-driven versus automatic learning},''
\newblock {\em Entropy}, vol. 21, no. 4, 2019.

\bibitem{Garcia2019ComputerTechniques}
José~Gabriel Garc{\'{i}}a, Adrián Colomer, Fernando L{\'{o}}pez-Mir, José~M.
  Mossi, and Valery Naranjo,
\newblock ``{Computer aid-system to identify the first stage of prostate cancer
  through deep-learning techniques},''
\newblock {\em European Signal Processing Conference}, pp. 1--5, 2019.

\bibitem{Nguyen2014ProstateNuclei}
Kien Nguyen, Anindya Sarkar, and Anil~K. Jain,
\newblock ``{Prostate cancer grading: Use of graph cut and spatial arrangement
  of nuclei},''
\newblock {\em IEEE Transactions on Medical Imaging}, 2014.

\bibitem{Nguyen2010AutomatedImages}
Kien Nguyen, Anil~K. Jain, and Ronald~L. Allen,
\newblock ``{Automated gland segmentation and classification for gleason
  grading of prostate tissue images},''
\newblock {\em International Conference on Pattern Recognition}, pp.
  1497--1500, 2010.

\bibitem{Nguyen2012StructureClassification}
Kien Nguyen, Anindya Sarkar, and Anil~K. Jain,
\newblock ``{Structure and context in prostatic gland segmentation and
  classification},''
\newblock {\em MICCAI 2012}, vol. 7510, pp. 115--123, 2012.

\bibitem{Garcia2018IdentificationTechniques}
Jose~Gabriel Garc{\'{i}}a, Adrián Colomer, Valery Naranjo, Francisco
  Pe{\~{n}}aranda, and M.~A. Sales,
\newblock ``{Identification of Individual Glandular Regions Using LCWT and
  Machine Learning Techniques},''
\newblock {\em IDEAL}, vol. 3, pp. 374--384, 2018.

\bibitem{Ronneberger2015U-net:Segmentation}
Olaf Ronneberger, Philipp Fischer, and Thomas Brox,
\newblock ``{U-net: Convolutional networks for biomedical image
  segmentation},''
\newblock {\em Lecture Notes in Computer Science (including subseries Lecture
  Notes in Artificial Intelligence and Lecture Notes in Bioinformatics)}, vol.
  9351, pp. 234--241, 2015.

\bibitem{He2016DeepRecognition}
Kaiming He, Xiangyu Zhang, Shaoqing Ren, and Jian Sun,
\newblock ``{Deep residual learning for image recognition},''
\newblock {\em Proceedings of the IEEE Computer Society Conference on Computer
  Vision and Pattern Recognition}, vol. 2016-Decem, pp. 770--778, 2016.

\bibitem{Ibtehaz2020MultiResUNet:Segmentation}
Nabil Ibtehaz and M.~Sohel Rahman,
\newblock ``{MultiResUNet: Rethinking the U-Net architecture for multimodal
  biomedical image segmentation},''
\newblock {\em Neural Networks}, vol. 121, pp. 74--87, 2020.

\bibitem{Szegedy2015GoingConvolutions}
Christian Szegedy, Wei Liu, Yangqing Jia, Pierre Sermanet, Scott Reed, Dragomir
  Anguelov, Dumitru Erhan, Vincent Vanhoucke, and Andrew Rabinovich,
\newblock ``{Going deeper with convolutions},''
\newblock {\em Proceedings of the IEEE Computer Society Conference on Computer
  Vision and Pattern Recognition}, vol. 07-12-June, pp. 1--9, 2015.

\bibitem{Kumar2017APathology}
Neeraj Kumar, Ruchika Verma, Sanuj Sharma, Surabhi Bhargava, Abhishek Vahadane,
  and Amit Sethi,
\newblock ``{A Dataset and a Technique for Generalized Nuclear Segmentation for
  Computational Pathology},''
\newblock {\em IEEE Transactions on Medical Imaging}, vol. 36, no. 7, pp.
  1550--1560, 2017.

\end{thebibliography}
\balance

\end{document}